\documentclass[twocolumn]{aa}
\usepackage{graphicx}
\usepackage{txfonts}
\usepackage{natbib}
\newcommand{\beq}   {\begin{equation}}
\newcommand{\eeq}   {\end{equation}}
\newcommand{\kms}   {km~s$^{-1}$}
\newcommand{\water}   {H$_2$O~}

\begin{document}
   \title{The Magnetic Field in the Star-forming Region Cepheus A}

   \titlerunning{Magnetic Fields in Cepheus A}

   \subtitle{from \water Maser Polarization Observations}

   \author{W.H.T. Vlemmings\inst{1}\and
           P.J. Diamond\inst{1}\and
           H.J. van Langevelde\inst{2,3}\and
	   J.M. Torrelles\inst{4}
          }

   \offprints{WV (wouter@jb.man.ac.uk)}

   \institute{Jodrell Bank Observatory, University of Manchester, Macclesfield,
                    Cheshire, SK11 9DL, England  
         \and
         Joint Institute for VLBI in Europe, Postbus 2, 
                7990~AA Dwingeloo, The Netherlands
	 \and
	Sterrewacht Leiden, Postbus 9513, 2300~RA Leiden, The Netherlands
	\and
	Instituto de Ciencias del Espacio (CSIS)-IEEC, C/ Gran Capit\'a, 2-4, 08034 Barcelona, Spain; on sabbatical leave at the UK Astronomy Technology Centre, Royal Observatory Edinburgh, U.K.
                          }

   \date{13-10-2005}

\abstract{ We present linear and circular polarization observations of
  the \water masers in 4 distinct regions spread over $1\times 2$
  arcseconds around the HW2 high-mass young stellar object in the
  Cepheus~A star-forming region. We find magnetic fields between
  100--500~mG in the central maser region, which has been argued to
  trace a circumstellar disk. The masers further from HW2 have field
  strengths between 30--100~mG. In all cases the magnetic field
  pressure is found to be similar to the dynamic pressure, indicating
  that the magnetic field is capable of controlling the 
  outflow dynamics around HW2. In addition to several \water maser
  complexes observed before, we also detect a new maser filament,
  $\simeq$~1$\arcsec$ ($\simeq$~690~AU) East of HW2, which we
  interpret as a shocked region between the HW2 outflow and the
  surrounding medium. We detect a linear polarization gradient along
  the filament as well as a reversal of the magnetic field
  direction. This is thought to mark the transition between the
  magnetic field associated with the outflow and that found in the
  surrounding molecular cloud. In addition to the magnetic field we
  determine several other physical properties of the maser region,
  including density and temperatures as well as the maser beaming
  angles.   }

   \maketitle

\section{Introduction}
\label{intro}


While the process of low-mass star-formation has been well studied,
high-mass star-formation is still poorly understood. Although several
theories propose the formation of high-mass stars from the merger of
several low-mass young stellar objects \citep[e.g.][]{Bonnell98} recent
studies and observations suggest that high-mass stars form, similar to
low-mass stars, through accretion from a circumstellar disk
\citep[e.g.][]{McKee03, Patel05, Jiang05}. In the prevailing picture
of low-mass star-formation out of dense molecular clouds, strong
magnetic fields support the clouds against a gravitational
collapse. When self-gravity overcomes the magnetic pressure in the
cloud core, the formation of protostars
ensues~\citep[e.g.][]{Shu87,Mouschovias99}. Additionally, magnetic
fields likely play an important role in many other stages of
star-formation, such as the formation of bi-polar outflows and a
circumstellar disk~\citep[e.g.][]{Akeson97}.
Thus,
accurate measurements of the magnetic field strength and structure in
the densest areas of star-forming regions (SFRs) are needed to
investigate the exact role of the magnetic field in both high- and
low-mass starformation \citep[see, e.g][]{Sarma01, Sarma02}.

Through polarization observations, masers are excellent probes of
magnetic field strength and structure in masing regions. For example,
polarimetric SiO, \water and OH maser observations in the envelopes of
evolved stars have revealed the strength and structure of the magnetic
fields during the end-stages of stellar
evolution~\citep[e.g][]{Kemball97,Etoka04,Vlemmings05b} and \water
maser polarization observations have provided stringent upper limits
of the magnetic field in the megamaser galaxy
NGC~4258~\citep{Modjaz05}. SFRs also show a rich variety of maser
species, including OH and \water. The OH masers are often found at
several hundred to thousands AU from the SFR cores where the density
$n_{\rm H_2}$ is less than a few times 10$^8$~cm$^{-3}$. Observations
of the Zeeman effect on OH masers have been used to determine the SFR
magnetic field in those regions \citep[e.g][]{Cohen90,
  Bartkiewicz05}. The \water maser emission in SFRs is often
associated with shocks created by the outflows of young stellar
objects (YSOs) or with a circumstellar disk (\citealp[~hereafter
  T96]{Torrelles96};~\citealp[~hereafter G03]{Gallimore03}). The
\water masers are excited in the dense parts of SFRs, with number
densities $n_{\rm H_2}$ between approximately 10$^8$ and
10$^{10}$~cm$^{-3}$ \citep{Elitzur89}. Because they are typically
small ($\sim$1~AU), have a narrow velocity width ($\sim$1~\kms) and
have a high brightness temperature $T_{\rm b}>10^9$~K
\citep[e.g.][]{Reid81}, \water masers can be used to examine the small
scale magnetic field strength and structure in dense parts of SFRs
with polarimetric very long baseline interferometry (VLBI)
observations. Previous VLBI observations have studied the linear
polarization of \water masers as tracer of the magnetic field
morphology in the SFRs W51~M \citep{Leppanen98}, Orion KL and W3~IRS~5
\citep{Imai03}. The circular polarization due to Zeeman splitting of
the 22~GHz \water masers was first observed by \citet{Fiebig89} with
the Effelsberg 100m telescope. These observations were confirmed with
VLBI by \citet{Sarma01}, who observed the \water maser circular
polarization in W3~IRS~5 with the Very Long Baseline Array (VLBA). At
lower spatial resolution, \citet{Sarma02}, also used the Very Large
Array (VLA) to determine magnetic field strengths in a number of SFRs
from \water maser observations.  Here we present VLBA linear and
circular polarization observations of the \water maser structures in
the SFR Cepheus~A~HW2.
 
\begin{figure}[t!]
   \resizebox{\hsize}{!}{\includegraphics{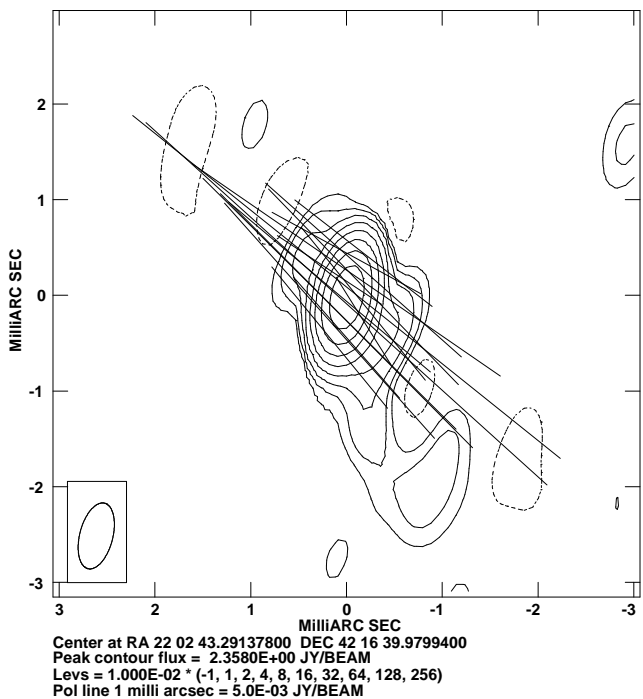}}
   \hfill
\caption[j2202]{Total intensity (I) map with polarization vectors of our polarization calibrator J2202+4216 (BL~Lac). The position angle ($\chi$) of the vectors has been rotated by 77$^\circ$ so that it corresponds to the VLBA calibration observation.}
\label{Fig:J2202}
\end{figure}

Cepheus~A is a high-mass SFR located at a distance of $\sim$725~pc
\citep{Johnson57}, which contains a large number of radio continuum
sources (HW sources; \citealp{Hughes84}). Additionally it exhibits
multi-polar outflows, NH$_3$ clouds, Herbig-Haro (HH) objects and
infrared sources and a complex structure of OH, \water and methanol
masers. The HW sources are compact HII regions that are thought to be
excited by a YSO either externally or embedded in the HII cloud itself
\citep{Cohen84, Garay96}. The brightest of these sources is HW2
\citep{Rodriguez94}, which is thought to contain the main exciting
source in the SFR. Surrounding it is a rich structure of \water masers
which has been studied in great detail \citep[e.g T96;
  G03;][~hereafter T98, T01a and T01b]{Torrelles98, Torrelles01b,
  Torrelles01a}. More \water maser structures are found in clusters
around other HW sources (HW3b and HW3d), $\simeq$~4-5$\arcsec$ south from
HW2 \citep[T98; ][]{Lada81, Cohen84, Rowland86}. The main, large
scale, \water maser structure in the direction of HW2 was interpreted
as tracing a 300~AU radius circumstellar disk perpendicular to the HW2
radio jet (T96). Recently, a flattened disk-like structure of dust and
molecular gas with radius $\simeq$~330~AU oriented perpendicular to
and spatially coincident with the HW2 radio jet has been reported
\citep{Patel05, Curiel05}.

Here we examine the polarization properties of the \water masers
around Cepheus~A~HW2 and determine the magnetic field strength and
structure. Additionally we describe the physical properties of the
\water maser regions and discuss the detection of a new \water maser
filament approximately 1\arcsec~East of the HW2 region.

The observations are described in \S~\ref{obs} and the results on the
maser morphology and polarization are presented in
\S~\ref{results}. The results are discussed in \S~\ref{disc}, where
intrinsic properties of the masing regions are derived. This is
followed by a summary and conclusions in \S~\ref{sum} and \S~\ref{concl}.
The analysis method and the \water maser models used are presented in
Appendix~\ref{method}.

\section{Observations}
\label{obs}

The observations were performed with the NRAO\footnote{The National
  Radio Astronomy Observatory (NRAO) is a facility of the National
  Science Foundation operated under cooperative agreement by
  Associated Universities, Inc.} VLBA on October 3 2004. The average
beam width is $\approx~0.5 \times 0.5$~mas at the frequency of the
$6_{16} - 5_{23}$ rotational transition of H$_2$O, 22.235080 GHz. We
used 4 baseband filters of 1~MHz width, which were overlapped to get a total
velocity coverage of $\approx~44$~\kms, covering most of the velocity
range of the H$_2$O masers around the mean velocity of the \water
masers of HW2 $V_{\rm lsr}=-11.7$~\kms\ (T96). Similar to the
observations in \citet{Vlemmings02}~(hereafter V02) of circumstellar
\water maser polarization, the data were correlated multiple times
with a correlator averaging time of 8~sec. The initial correlation was
performed with modest spectral resolution (128 channels; $7.8$~kHz$ =
0.1$~\kms), which enabled us to generate all 4 polarization
combinations (RR, LL, RL and LR). Two additional correlator runs were
performed with high spectral resolution (512 channels; $1.95$~kHz$ =
0.027$~\kms), which therefore only contained the two polarization
combinations RR and LL, to be able to detect the signature of the
H$_2$O Zeeman splitting across the entire velocity range. The
observations on Cepheus A~HW2 were interspersed with 15 minute
observations of the polarization calibrator J2202+4216
(BL~Lac). Including scans on the phase calibrators (3C345 and 3C454.3)
the total observation time was 8 hours.

\subsection{Calibration}

The data analysis path is described in detail in V02. It follows the
method of \citet{Kemball95} and was performed in the Astronomical
Image Processing Software package (AIPS). The calibration steps were
performed on the data-set with modest spectral resolution. Delay,
phase and bandpass calibration were performed on 3C345, 3C454.3 and
J2202+4216.  Polarization calibration was performed on the
polarization calibrator J2202+4216 (Fig.\ref{Fig:J2202}). Fringe
fitting and self-calibration were performed on a strong ($\sim$80~Jy beam$^{-1}$)
maser feature (at $V_{\rm lsr}=-15.72$~\kms). The calibration
solutions were then copied and applied to the high spectral resolution
data-set. Finally, corrections were made for instrumental feed
polarization using a range of frequency channels on the maser source,
in which the expected frequency averaged linear polarization is close
to zero. In order to make a comparison with previous results we have
used the AIPS task FRMAP in an attempt to determine the position of
the reference feature before any self-calibration or fringe
fitting. Though an exact position determination was impossible, we
found it to be within $\sim$25~mas of our pointing position
($\alpha(J2000)=22^h56^m17^s\!.977$ and
$\delta(J2000)=+62^{\circ}01^{\arcmin}49^{\arcsec}\!.419$), which was
the brightest maser feature of the maser region R4 from G03.

An initial image cube with low resolution ($2048\times2048$ pixels of
1~mas) was created from the modest spectral resolution data set using
the AIPS task IMAGR. In this cube a search was performed for maser
features and 4 distinct regions with maser emission were detected
(further labeled I through IV; shown in Fig.~\ref{Fig:FULL}). For
these fields, typically $\sim$100$\times$100~mas in size, IMAGR was
used to create high spatial resolution ($1024\times1024$ pixels of
0.09~mas) Stokes I, Q and U image cubes from the modest spectral
resolution data set. Stokes I and V cubes for the same regions were
created from the high spectral resolution data set. In the high
spectral resolution total intensity channel maps, the noise ranges
from $\approx$15~mJy in the channels with weak maser features, to
$\approx$35~mJy when dominated by dynamic range effects in the
channels with the strongest maser features. In the circular
polarization polarization maps the rms noise is $\approx$15~mJy. In
the lower resolution Stokes Q and U maps the rms noise is $\approx~10$~mJy.

Unfortunately, we found that in a small range of frequency channels
where a higher frequency band overlaps the neighboring lower band,
cross-talk between the sub-bands resulted in unreliable
calibration. Although we were able to image the masers in those
channels ($V_{\rm lsr}$ between $-12.8$~\kms~and $-13.5$~\kms~as well
as between $-1.2$~\kms~and $-2.5$~\kms) they were not included in our
polarization analysis as the calibration accuracy was insufficient.

To calibrate the polarization angle $\chi = 1/2\times{\rm atan}(U/Q)$ of
the resulting maps, the polarization calibrator J2202+4216 was mapped
using the full 4~MHz bandwidth. The resulting map with
polarization vectors is shown in Fig.~\ref{Fig:J2202}. The
polarization vectors were rotated to match the polarization angle of
J2202+4216 determined in the VLBA polarization calibration
observations \footnote{http://www.aoc.nrao.edu/$\sim$smyers/calibration/}.
As our observations were made exactly between two of the calibration
observations on September 19 and October 17 2004 where the polarization
angle of J2202+4216 changed from 41$^\circ$ to 57$^\circ$, we use the
average of 49$^\circ$. Thus, we estimate our polarization angles to
contain a possible systematic error of $\sim$8$^\circ$.

\subsection{Cepheus~A~HW2}

\begin{figure*}[t!]
   \resizebox{\hsize}{!}{\includegraphics{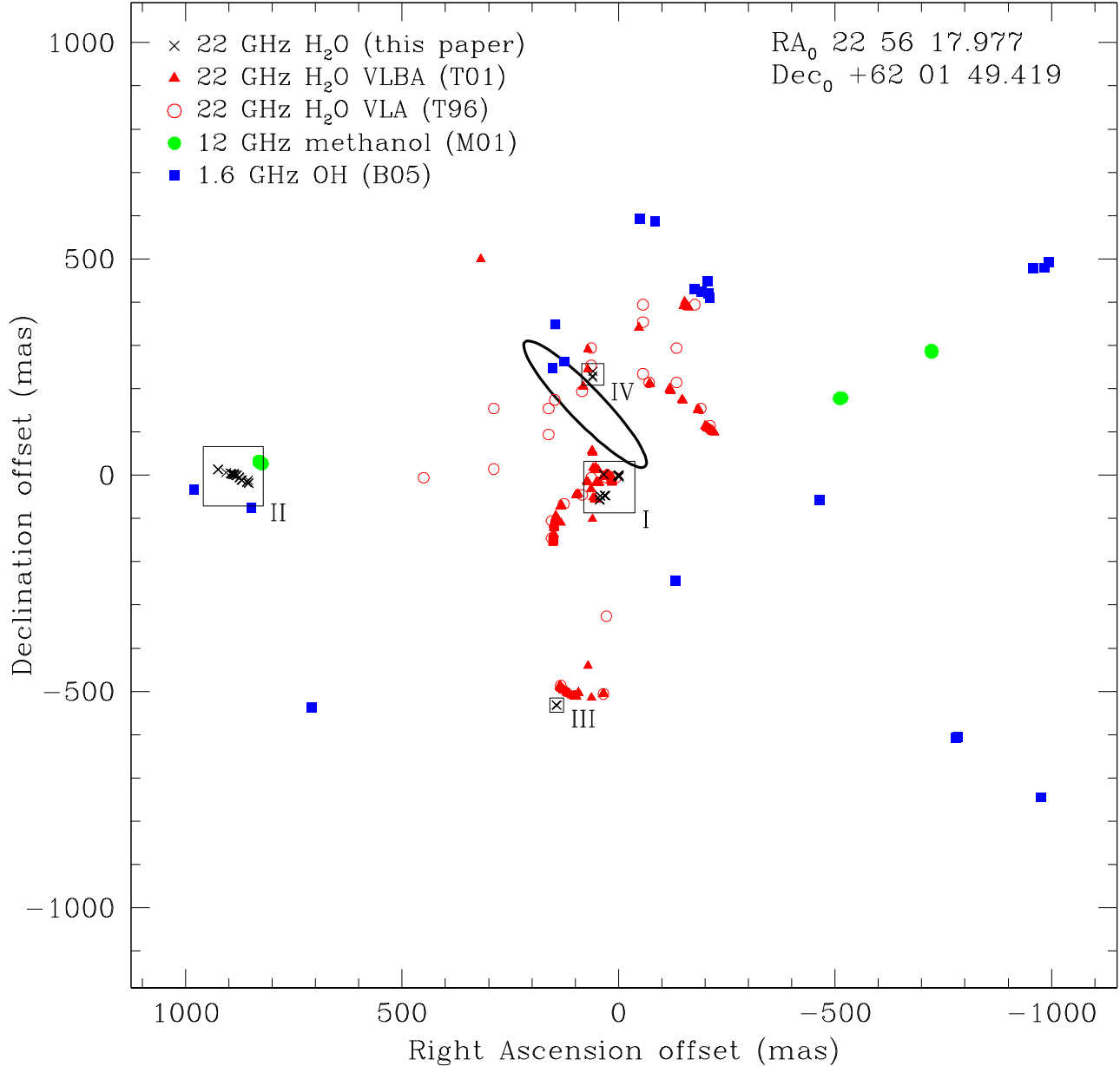}}
   \hfill \caption[fullview]{The Cepheus A~HW2 region with related
   maser features. The crosses indicate the \water maser features
   from our observations with the boxes labeled I through IV
   the fields in which the masers where detected. The other \water
   maser positions are from the VLA and VLBA observations in T96 and
   T01 respectively. The solid dots are the positions of the 12~GHz
   methanol masers from \citet{Minier01}~(M01) The solid squares are the 1665
   and 1667~MHz OH masers observed by \citet{Bartkiewicz05}~(B05). The
   ellipse denotes the position and shape of the HW2 continuum
   emission at 1.3~cm (T96).}
   \label{Fig:FULL} \end{figure*}

We detected 4 distinct regions of \water maser emission between
$V_{\rm lsr}=-22.5$ and $0.5$~\kms. We did not detect any of the maser
features with positive velocity from T98 and T01a to a limit of
$\approx$45~mJy. In Fig.~\ref{Fig:FULL} we show a $2.3\arcsec\times
2.3\arcsec$ area around HW2 in which the fields where \water maser
emission was detected are marked. We also indicate the continuum
source HW2 (T96) and the location of previously detected \water maser
not visible in our observations. Additionally, the location of OH
masers \citep{Bartkiewicz05} and 12~GHz methanol masers
\citep{Minier01} are plotted.  All offset positions in this paper are
given with respect to reference maser feature position at $V_{\rm
  lsr}=-15.72$~\kms\ which was earlier found to be within 25~mas of our
pointing center.
The accuracy of each individual maser feature
position can be estimated by ${\rm Beamsize}/{\rm SNR}$, which is
typically better than $\sim$0.005~mas. In our polarization analysis we
only considered maser features with intensities $>1$~Jy.

\begin{figure*}[t!]
\vskip15truecm
\caption[Mosaic]{A close-up view of the 4 fields in which we detected \water maser features. The octagonal symbols are the identified maser features scaled logarithmically according to their peak flux density. The maser velocity is indicated by color, note that the color scale is different for the 4 fields. A 10~Jy beam$^{-1}¶$ symbol is plotted for illustration in the lower left corner of Field IV. The linear polarization vectors, scaled logarithmically according to polarization fraction $P_l$, are over-plotted. For the maser features where the Zeeman splitting was detected the magnetic field strength is indicated in mG.}
\label{Fig:MOSAIC}
\end{figure*}

\begin{figure*}[t!]
   \resizebox{\hsize}{!}{\includegraphics{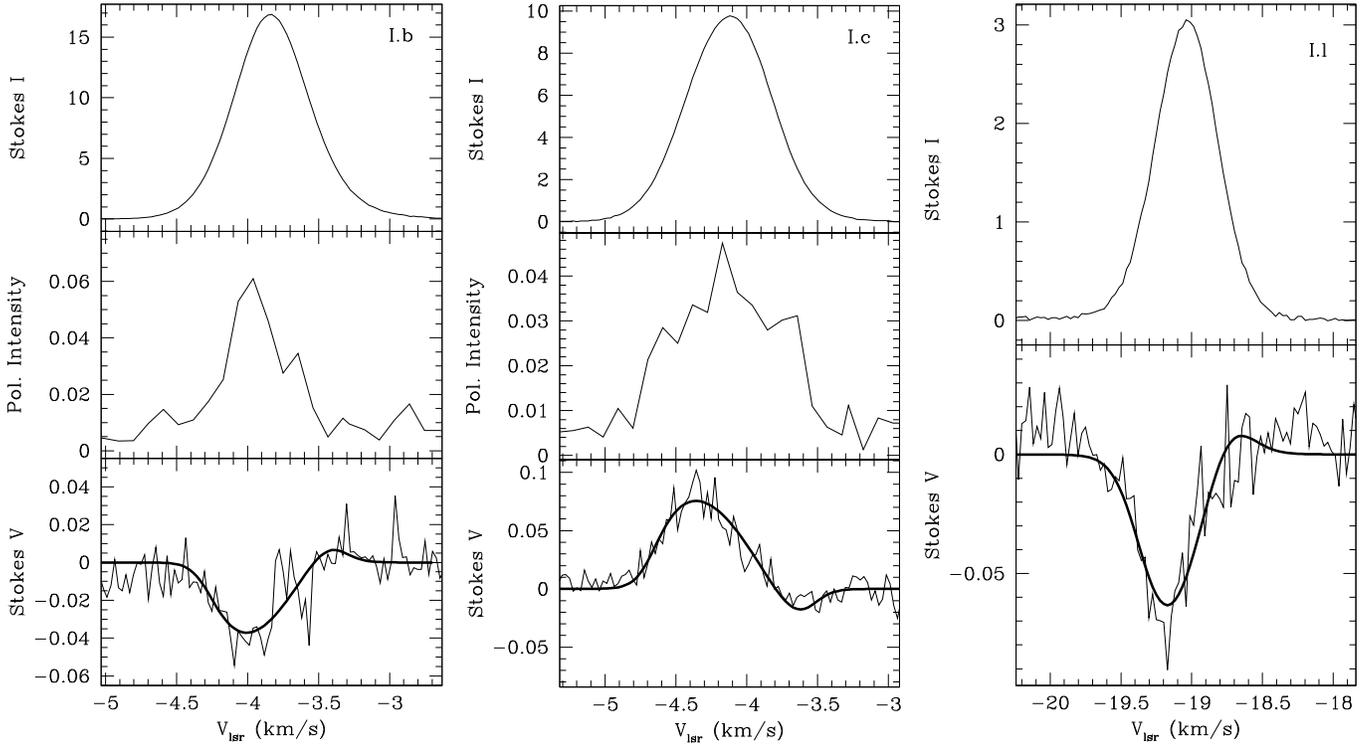}}
   \hfill \caption[VspecField1]{Total power (I) and V-spectra for selected maser features of Field I. Additionally, the linear polarized flux density, $\sqrt(Q^2+U^2)$, is shown when detected. The flux densities are given in Jy~beam$^{-1}$. The thick solid line in the bottom panel shows the best non-LTE model fit to the circular polarization V. The V-spectrum is adjusted by removing a scaled down version of the total power spectrum as indicated in Appendix~\ref{method}.}  
\label{Fig:V1} 
\end{figure*}

\section{Results}
\label{results}

\begin{table*}[t!]
\caption{Results}
\begin{tabular}{|l|c|c|c|c|c|c|c|c||c|c|c|}
\hline
Feature  & RA & Dec & Peak Flux & $V_{\rm lsr}$ & $\Delta
v_{\rm L}$ & P$_{\rm l}$ & $\langle\chi\rangle$ & P$_{\rm V}$ & B~$_{||}~^a$ & $\Delta v_{\rm th}~^a$ & $\log(T_{\rm b}\Delta\Omega)~^a$ \\
 & offset & offset & Density (I) & & & & & & & & \\
 & (mas) & (mas) & (Jy beam$^{-1}$) & \kms & \kms & (\%) & ($^\circ$) & $(\times10^{-3})$ & ${\rm (mG)}$ & \kms & \\
\hline
\hline
I.a$^*$ & 44.213 & -57.325 & 10.32 & -1.86  & 0.53 & - & - & - & - & - & - \\  
I.b & 43.289 & -56.585 & 16.79 & -3.83  & 0.59 & $0.28 \pm 0.12$ & $13 \pm 8 $ & 2.6 & $62\pm12$ & 1.8 & 10.5 \\
I.c & 42.750 & -56.510 & 9.83  & -4.12  & 0.70 & $0.44 \pm 0.06$ & $35 \pm 8 $ & 9.5 & $-290\pm47$ & 2.0 & 10.7 \\
I.d$^{**}$ & 42.290 & -47.644 & 19.09 & -4.28  & 0.58 & $0.78 \pm 0.06$ & $-46 \pm 3 $ & 12.7 & $-279\pm69$ & - & - \\
I.e & 42.229 & -48.655 & 75.18 & -3.96  & 0.78 & $0.64 \pm 0.04$ & $-47 \pm 8 $ & 10.3 & $205\pm40$ & 1.0 & 10.0\\
I.f$^*$ & 34.428 &   1.488 & 33.35 & -13.04 & 0.48 & - & - & - & - & - & - \\
I.g$^*$ & 33.811 &   1.977 & 27.49 & -12.97 & 0.53 & - & - & - & - & - & - \\
I.h$^*$ & 32.322 & -46.771 & 87.44 & -1.94  & 0.62 & - & - & - & - & - & - \\
I.i$^*$ & 31.031 & -47.121 & 12.81 & -1.68  & 1.01 & - & - & - & - & - & - \\
I.j$^*$ & 29.643 & -47.557 & 8.49  & -1.20  & 0.94 & - & - & - & - & - & - \\
I.k &  1.963 &  -1.027 & 3.48  & -18.83 & 0.57 & $<0.86$ & - & - & $<206$ & - & - \\
I.l &  1.040 &  -2.099 & 3.04  & -19.04 & 0.51 & $<0.99$ & - & 23.1 & $527\pm109$ & 2.0 & 9.9 \\
I.m &  0.082 &  -3.303 & 11.47 & -19.44 & 0.50 & $<0.26$ & - & 6.8 & $135\pm26$ & 1.7 & 10.3 \\
I.n &  0.000 &   0.000 & 78.94 & -15.72 & 0.52 & $<0.04$ & - & 7.2 & $148\pm34$ & 1.7 & 10.4 \\
I.o & -0.820 &  -4.152 & 4.02  & -19.28 & 0.59 & $1.17 \pm 0.19$ & $8 \pm 9$ & - & $<256$& - & -  \\
I.p & -0.871 &  -0.183 & 61.63 & -15.89 & 0.51 & $0.42 \pm 0.02$ & $58 \pm 2$ & 6.9 & $-150\pm42$ & 2.0 & 9.8 \\
I.q$^{**}$& -1.486 &  -1.359 & 46.82 & -16.78 & 0.68 & $0.59 \pm 0.26$ & $39 \pm 2$ & 6.8 & $-203\pm71$ & - & - \\
\hline
II.a & 925.674 & 13.463 &  18.07 & -14.91 & 0.51 & $0.33 \pm 0.09$ & $-41\pm5$ & -  & $<22$ & - & - \\  
II.b & 924.445 & 13.075 &  44.81 & -14.93 & 0.49 & $<0.07$ & - & 3.5 & $-54\pm9$ & 1.5 & 9.8 \\
II.c & 906.332 &  6.070 &  2.76  & -14.25 & 0.48 & $<1.09$ & - & & $<140$ & - & - \\
II.d & 896.763 &  3.990 &  15.24 & -14.06 & 0.51 & $0.52 \pm 0.01$ & $9 \pm 7 $ & - & $<34$ & - & - \\
II.e & 894.765 &  3.352 &  21.08 & -13.99 & 0.59 & $0.37 \pm 0.05$ & $-12 \pm 10$ & - & $<28$ & - & - \\
II.f & 892.838 &  2.587 &  24.33 & -14.09 & 0.46 & $0.21 \pm 0.05$ & $38 \pm 12 $ & - & $<30$ & - & - \\
II.g & 890.204 &  1.947 &  3.79  & -13.99 & 0.33 & $<0.79$ & - & - & $<88$ & - & - \\
II.h & 888.068 &  3.166 &  3.20  & -13.55 & 0.40 & $3.44 \pm 1.13$ & $-78 \pm 4  $ & - & $<125$& - & -  \\
II.i & 887.387 &  1.437 &  33.64 & -13.70 & 0.50 & $0.15 \pm 0.02$ & $-50 \pm 16$ & - & $<15$ & - & -  \\
II.j & 885.188 &  0.222 &  8.03  & -13.60 & 0.47 & $1.05 \pm 0.10$ & $6 \pm 4  $ & - & $<59$ & - & - \\
II.k$^*$ & 882.617 & -0.579 &  11.25 & -13.47 & 0.53 & - & - & - & - & - & - \\
II.l$^*$ & 881.495 & -0.880 &  6.95  & -13.31 & 0.47 & - & - & - & - & - & - \\
II.m$^*$ & 880.610 &  0.001 &  5.86  & -13.26 & 0.46 & - & - & - & - & - & - \\
II.n$^*$ & 876.069 & -4.921 &  5.25  & -12.81 & 0.40 & - & - & - & - & - & - \\
II.o$^*$ & 875.052 & -5.871 &  5.03  & -12.81 & 0.44 & - & - & - & - & - & - \\
II.p & 870.917 &-11.180 &  2.48  & -12.57 & 0.47 & $<1.21$ & - & - & $<190$ & - & - \\
II.q & 869.659 &-11.433 &  5.84  & -12.65 & 0.50 & $<0.51$ & - & 3.5 & $66\pm33$ & 1.8 & 10.2 \\
II.r$^*$ & 858.940 &-15.758 &  7.85  & -12.91 & 0.41 & - & - & - & - & - & - \\
II.s$^*$ & 857.685 &-16.246 &  8.92  & -12.91 & 0.40 & - & - & - & - & - & - \\
II.t & 855.098 &-11.502 &  5.21  & -14.51 & 0.55 & $0.41 \pm 0.07$ & $86 \pm 19 $ & - & $<127$ & - & - \\
II.u$^*$ & 854.380 &-17.668 &  56.09 & -12.83 & 0.42 & - & - & - & - & - & - \\
II.v & 853.907 &-18.350 &  23.43 & -12.78 & 0.41 & $0.34 \pm 0.09$ & $84 \pm 7  $ & 5.7 & $69\pm11$ & 1.1 & 9.7 \\
\hline
III.a & 143.668 & -530.929 & 21.04 & -8.25 & 0.58 & $10.8\pm0.9$ & $66\pm1$ & 1.8 & $33\pm10$ & 1.2 & 10.7 \\
III.b & 142.096 & -532.309 & 6.02 & -8.52 & 0.57 & $5.0\pm0.8$ & $62\pm2$ & 6.9 & $128\pm36$ & 1.3 & 10.5 \\  
\hline
IV.a & 60.131 & 229.423 & 1.49 & -21.28 & 0.62 & $<2.01$ & - & -  & $<520$ & - & - \\  
IV.b & 60.025 & 226.307 & 2.28 & -21.07 & 0.65 & $1.35\pm0.24$ & $64\pm2$ & -  & $<356$ & - & - \\  
IV.c & 58.833 & 240.075 & 1.38 & -20.75 & 0.86 & $<2.17$ & - & -  & $<779$ & - & - \\  
\hline
\multicolumn{11}{l}{$^a$ Best fit results for the magnetic field strength $B_{||}$ along the line of sight (mG), intrinsic maser thermal width $\Delta v_{\rm th}$ (\kms)} \\
\multicolumn{11}{l}{and emerging brightness temperature $T_b\Delta\Omega$ (K~sr) derived as described in Appendix~\ref{method}} \\
\multicolumn{11}{l}{$^*$ suffer from interference (see text)} \\
\multicolumn{11}{l}{$^{**}$ no direct fit possible (see Appendix~\ref{method})} \\
\end{tabular}
\label{Table:results}
\end{table*}

\subsection{Distribution of the Maser Features}

In Fig.~\ref{Fig:MOSAIC} we show the 4 fields in which maser features
stronger than 1~Jy beam$^{-1}$ were identified. The hexagonal symbols denoting the
maser features are scaled logarithmically by their flux density level. We identified
54 maser features although 14 of those had a velocity located in the
ranges that suffered from interference as discussed above. The maser
features are listed in Table~\ref{Table:results} with their positional
off-set from the reference maser position, peak flux density, radial velocity
$V_{\rm lsr}$ and full width half maximum (FWHM) $\Delta v_L$. The
positions were determined in the frequency channel containing the peak Stokes I
emission using the AIPS task JMFIT. The masers in Field I, III and IV
were seen previously (T01b) while Field II contains a newly detected
linear maser structure approximate 1\arcsec~East of HW2 (assuming a
distance of 725~pc the masers are located $\sim$690~AU from HW2). The
masers in Field I are identified as the masers seen in the region
labelled R4 of T01b observed with similar lsr velocity. This are also the masers that were hypothesized
to belong to a rotating disk in G03 and are found over a large
velocity range ($V_{\rm lsr}$ between $-20$ and 0~\kms). As seen in
Fig.~\ref{Fig:FULL}, the maser structure in Field II is located close
to the brightest of the 12~GHz methanol maser features detected by
\citet{Minier01}. However, the methanol masers at $V_{\rm
lsr}=-4.2$~\kms\ are significantly red-shifted with respect to the
\water maser structure, which has an average $V_{\rm
lsr}\sim -13.7$~\kms. The 2 maser features in Field III at $V_{\rm
lsr}\approx -8.5$~\kms, correspond to a small part of the extended
maser arc in R5 of T01b. This arc was identified to belong to a
spherical shell around an embedded YSO \citep{Curiel02}. The fairly weak
masers of Field IV at $V_{\rm lsr}\approx -21$~\kms\ are located closest
to HW2 and likely correspond to a few isolated features detected in
T96. We did not detect any of the masers from the arc-like structures
in R1, R2 and R3 of T01b. The total extent of the region in which we
detected maser emission is $\sim 950 \times 790$~mas, corresponding to $690
\times 575$~AU.

\subsection{Circular Polarization}

\begin{figure*}[t!]
   \resizebox{\hsize}{!}{\includegraphics{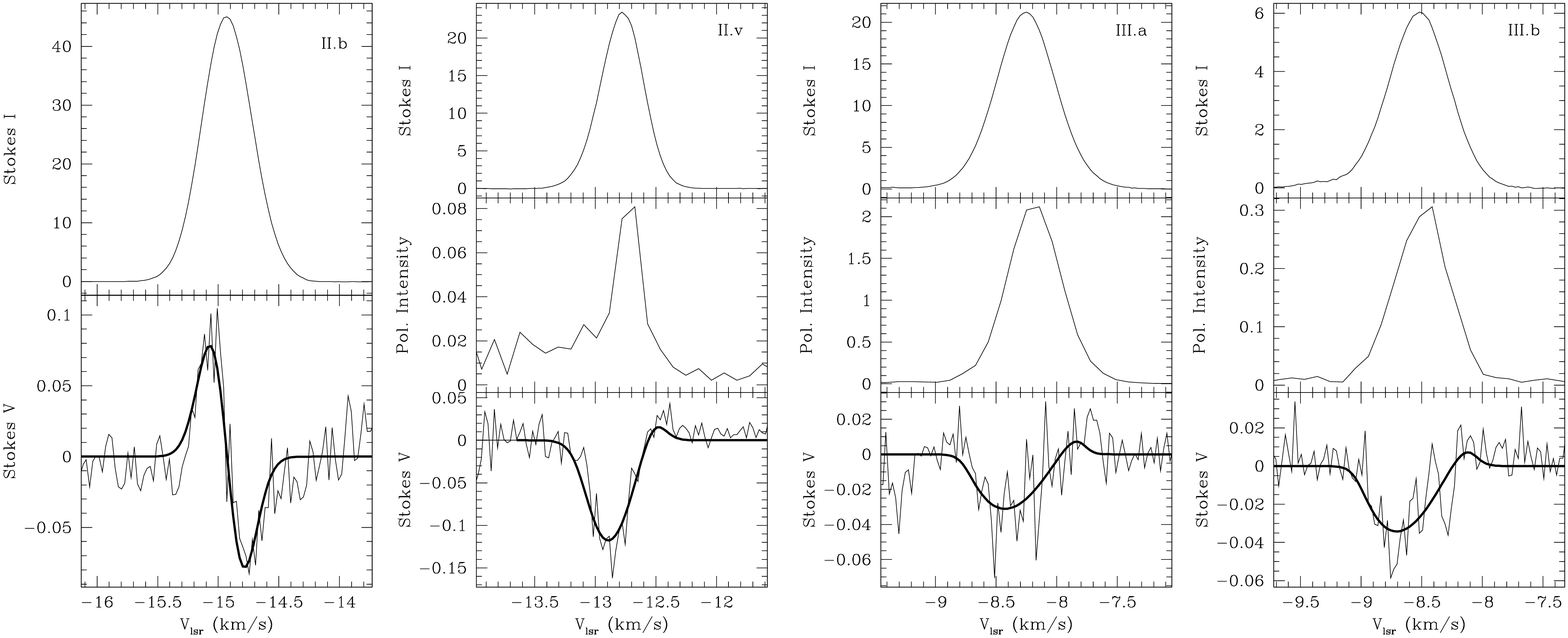}}
   \hfill
\caption[VspecField2]{Similar to Fig.~\ref{Fig:V1} for selected \water masers in Field II and both features in Field III.}
\label{Fig:V2}
\end{figure*}

Circular polarization between 0.018--2.31\% was detected in 14 of the
40 maser features that did not suffer from the frequency band overlap
interference. Features that were not analyzed due to the interference
are marked in Table~\ref{Table:results}. This table also shows the
circular polarization fraction $P_V$ as well as the magnetic field
strengths along the line of sight with $1\sigma$ errors or $3\sigma$
upper limits determined by comparing the line width and circular
polarization with models of non-LTE radiative transfer in the
magnetized \water molecules (Appendix~\ref{method}). As the $1\sigma$
errors include both the formal fitting uncertainties as well as the
contribution of the error in the model $\Delta v_{\rm th}$ (thermal
line width) and $T_{\rm b}\Delta\Omega$ (emerging maser brightness
temperature in K~sr), the magnetic field strength can occasionally be
$<3\sigma$, even though the circular polarization signal has a SNR
higher than 3. The table also includes the best fit model values for
$\Delta v_{\rm th}$ and $T_{\rm b}\Delta\Omega$, where the emerging
brightness temperature has been scaled with maser decay and
cross-relaxation rate as described in Appendix~\ref{method}. The
errors on these are estimated there to be $0.3$~\kms\ in $\Delta v_{\rm
  th}$ and $0.4$ on ${\rm log}(T_{\rm b}\Delta\Omega)$. As the lack of
circular polarization introduces an additional free parameter in the
model fitting, significantly increasing the $\Delta v_{\rm th}$ and
$T_{\rm b}\Delta\Omega$ errors, we do not fit for maser features that
do not show circular polarization. The magnetic field strength ranges
from several tens of mG in Fields II and III to several hundred mG in
Field I and is seen to switch direction on small scales in both Field
I and II. Note that a positive magnetic field values indicates a field
pointing away from the the observer. Total intensity (I) and circular
polarization (V) spectra of several of the maser features are shown in
Figs.~\ref{Fig:V1} and \ref{Fig:V2}. The spectra include the best fit
model for the circular polarization.

\begin{figure*}
   \resizebox{\hsize}{!}{\includegraphics{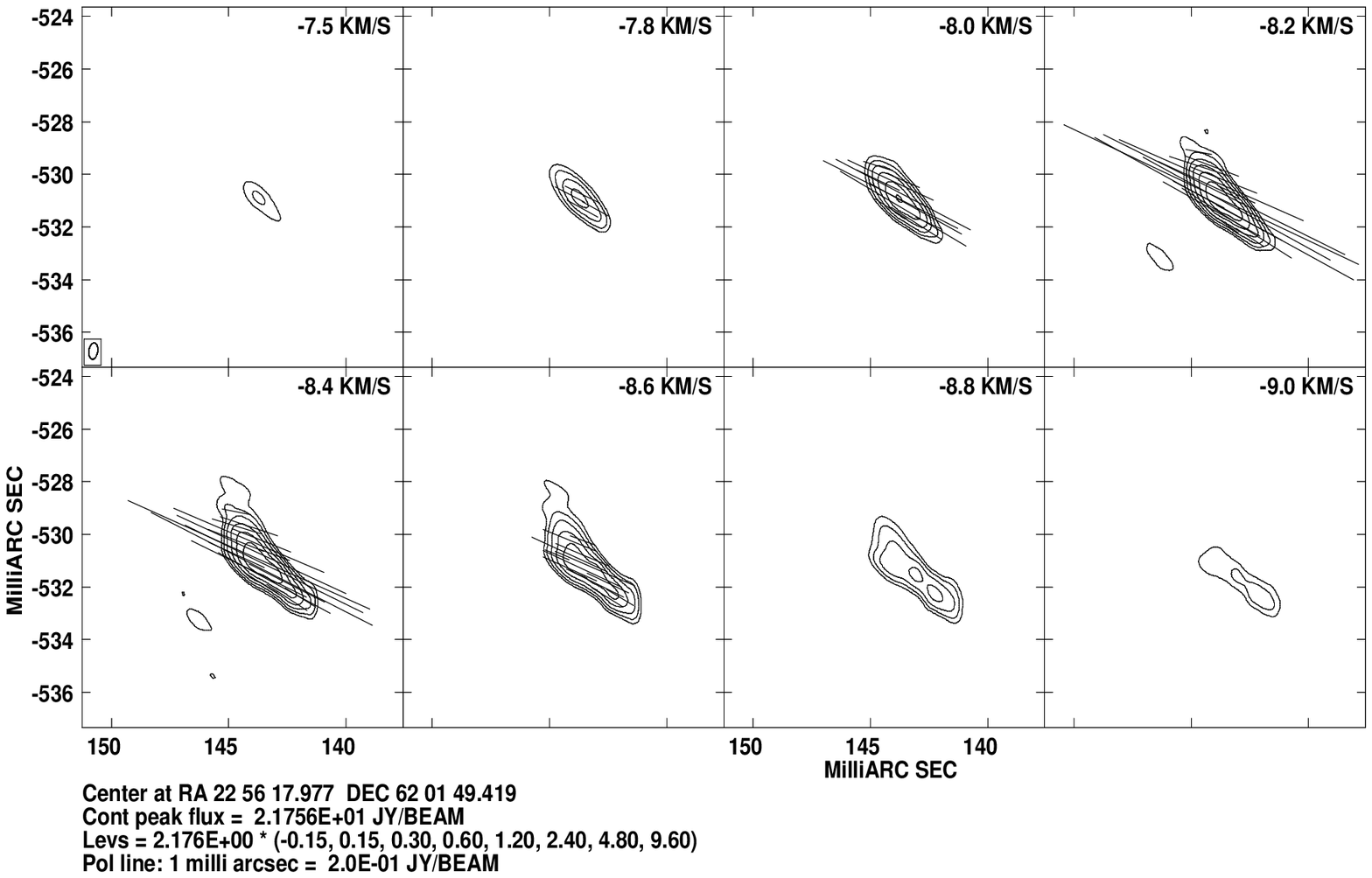}}
   \hfill
\caption[LinpolField3]{Channel maps of linear polarization of the elongated \water maser feature of Field III which has the highest linear polarization fraction. The bars show the strength and orientation of the polarization vectors.}
\label{Fig:LP3}
\end{figure*}

\subsection{Linear Polarization}

In addition to the circular polarization, we detected linear
polarization in approximately 50\% of our maser features. The
fractional linear polarization $P_l$ is given in
Table~\ref{Table:results}. Figs.~\ref{Fig:V1} and \ref{Fig:V2} also show several linear polarization
spectra. Table~\ref{Table:results} lists the weighted mean polarization
vector position angle $\langle\chi\rangle$ determined over the
maser FWHM for the linearly polarized maser features with
corresponding rms error. The weights are determined using the formal
errors on $\chi$ due to thermal noise, which are given by
$\sigma_\chi=0.5~\sigma_P/P \times 180^\circ/\pi$
\citep{Wardle74}. Here $P$ and $\sigma_P$ are the polarization
intensity and corresponding rms error respectively.
Fig.~\ref{Fig:MOSAIC} shows the linear polarization vectors scaled
logarithmically according to fractional polarization.

The strongest linear polarization ($\sim$11\%) was detected on the
brightest maser feature in Field III, but on average $P_l\sim$0.5\%.  In
Fig.~\ref{Fig:LP3} we show a channel map of the 2 maser features
detected in Field III including their polarization vectors. We do not
find any relation between maser brightness and fractional linear
polarization.

\section{Discussion}
\label{disc}

\begin{figure}
   \resizebox{\hsize}{!}{\includegraphics{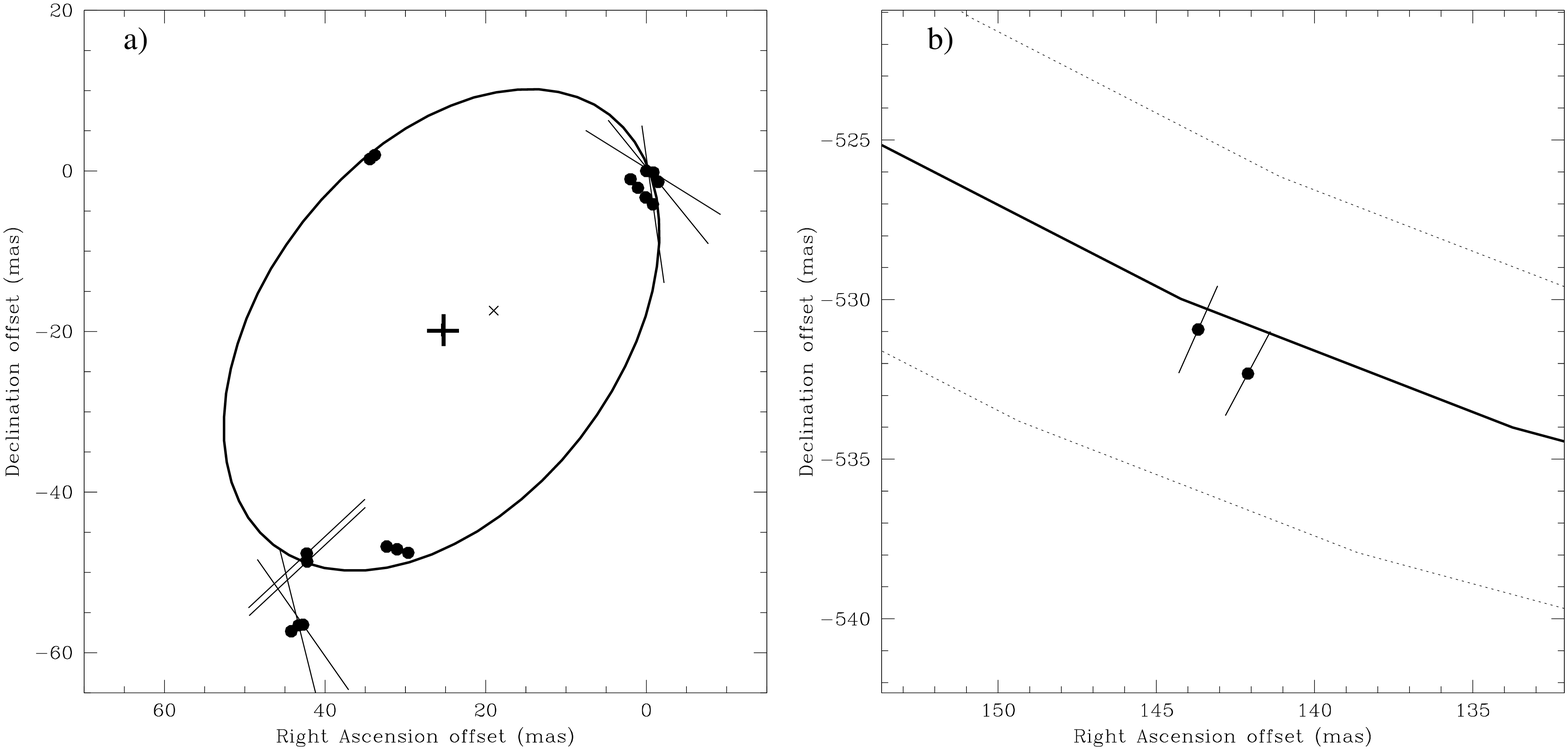}}
   \hfill \caption[shell]{a) The masers of Field I with the disk-model
   of G03. We fitted a Right Ascension and Declination offset of the
   disk center
   (denoted by the plus sign) as well as a Radius ($R_d=34 \pm
   4$~mas). The inclination angle of $50^\circ$ and P.A. of
   $142^\circ$ were taken from G03. The diagonal cross is the disk center
   position determined by G03 which has an error of $\approx$27~mas in
   each coordinate. The vectors on the maser features are the polarization vectors,
   which for most of the features is expected to be parallel to the
   magnetic field direction (see \S~\ref{disclp}). b) The masers of Field III with the
   expanding
   shell model of T01a. The solid line is the
   shell using the updated proper motion and expansion velocity
   parameters of G03. The dashed lines are the
   $3\sigma$ confidence interval. The vector on the maser features
   indicate the magnetic field direction (see \S~\ref{disclp}).}  \label{Fig:shell}
   \end{figure}

Before discussing the polarization results we first determine several
intrinsic properties of the masers that are needed for the further
analysis of the linear and circular polarization. We also discuss the
maser morphology in the 4 fields.

\subsection{Intrinsic Thermal Width, Brightness Temperatures and Maser Beaming}

As the model results give the intrinsic thermal width $\Delta v_{\rm th}$ in
the maser region, we can use it to estimate the temperature.  Although
the error on $\Delta v_{\rm th}$ is relatively large due to velocity
gradients along the maser \citep{Vlemmings05a}, we
find that on average, $\Delta v_{\rm th}$, and correspondingly the
temperature, is greater in Field I than in the outlying fields II and
III. While in Field I $T\sim$1150~K, in Field II and III the corresponding
temperature is closer to 750~K.  These temperatures are an indication
that the masers originate in a C-type (non-dissociative) shock instead
of a J-type (dissociative) shock. In the latter, the \water masers have
been found to originate in a relatively narrow range of temperatures
near 400~K (with 500~K as a conservative upper bound) at which
hydrogen molecules recombine. In contrast, in C-type shocks, the
\water masers can occur in gas with temperatures up to $\sim$3000~K
provided the shock velocity $v_s>10$~\kms\ \citep{Kaufman96}. The
C-shock origin of the masers in Field I is in agreement with the model
in G03 where the masers originate in a C-shock expanding though a
circumstellar disk.

In addition to $\Delta v_{\rm th}$ the models also provide an estimate of
the emerging brightness temperature $T_{\rm b}\Delta\Omega$. This can
be compared with the values determined from the measurements of the
maser flux density and feature sizes. We find that in Field I the
majority of the maser features are unresolved. Taking 0.4~mas as the
typical size of a \water maser feature, we derive a  the
brightness temperature of $T_{\rm b}\approx$1.4$\times10^{11}$~K for a
feature of 10~Jy beam$^{-1}$. Thus, our strongest maser feature in
Field I has $T_{\rm b}\approx$1.1$\times10^{12}$~K. In the other
fields, several of the masers are marginally resolved, with typical
feature sizes of $\sim$0.6~mas, corresponding to
$\sim$7.5$\times10^{12}$~cm. This implies, for the strongest 54~Jy
beam$^{-1}$ maser feature in those regions, $T_{\rm
  b}\approx$3.4$\times10^{11}$~K. Comparing these values with the
emerging brightness temperatures $T_{\rm b}\Delta\Omega$ from our
models yields an estimate for the beaming solid angle
$\Delta\Omega$. In Field I, with an average $\langle T_{\rm
  b}\Delta\Omega\rangle\approx$1.8$\times10^{10}$ we find, for the
maser features with circular polarization,
$\Delta\Omega\approx$7$\times10^{-3}$--3$\times10^{-1}$~sr.  However,
as the features are unresolved the beaming angle may be
overestimated.  The masers in Field II show a similar range of beaming
angle, with $\Delta\Omega\approx$2$\times10^{-2}$--4$\times10^{-1}$~sr
while the beaming of the maser in Field III is much less pronounced,
as $\Delta\Omega\approx$0.5. In a tubular geometry the maser beaming
$\Delta\Omega\approx~(d/l)^2$, where $d$ and $l$ are the transverse
size and length of the tube respectively, this implies that, assuming
$d$ is approximately the size of the maser features, the maser lengths
are $\sim$1--6$\times10^{13}$~cm. In Field III, the beaming angle is
similar to what is expected for a spherical maser that is approaching
saturation \citep{Elitzur94}.

We now compare our measured and derived maser brightness temperatures
with the maser brightness temperature $T_{\rm S}$ at the onset of
saturation when the ratio between maser rate of stimulated emission ($R_m$) and the maser decay rate ($\Gamma$), $R_m/\Gamma\approx$1. Using the expression from \citet{Reid88}: 
\begin{equation}
T_{\rm S}\Delta\Omega = h\nu\Gamma 4\pi / 2k_B A, 
\end{equation} 
where $h$ is the Planck constant and $k_B$ the Boltzmann constant,
$\nu$ is the maser frequency and $A=2\times10^{-9}$~s$^{-1}$ is the
22~GHz \water maser spontaneous emission rate \citep{Goldreich72}. For
$\Gamma=1$~s$^{-1}$ we thus find $T_{\rm
S}\Delta\Omega=3.4\times10^9$~K~sr. Nearly full saturation is reached when
$R_m/\Gamma\approx$100, for $T_{\rm
b}\Delta\Omega\approx3.4\times10^{11}$~K~sr. This indicates that in
Field II the masers are likely mostly unsaturated, while those in Field I are
in the onset of the saturation regime. In Field III the masers are almost fully
saturated, which is consistent with their strong linear
polarization (see Appendix~\ref{method} below). When saturated, the maser radiative transfer equation can
be approximated by $T_{\rm B}/T_{\rm S} \approx~g_0l$ where $g_0$ is
the maser gain at line center for the unsaturated regime. For the
masers in Field III we then find $g_0l\approx$8, indicating that for
$l\approx$1.5$\times10^{13}$~cm estimated from the beaming angle,
$g_0\approx$5$\times10^{-13}$~cm$^{-1}$.

\subsection{\water Maser Morphology}

\begin{figure}[ht!]
   \resizebox{\hsize}{!}{\includegraphics{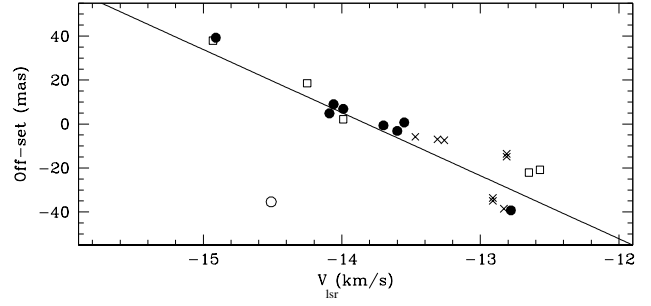}}
   \hfill
\caption[vel2]{The velocity of the \water maser features making up the
  filament detected in Field II vs. angular off-set from the center of
  the filament. The solid and open circles are the maser features that
  have detected linear polarization while the open squares are the
  features for which we determined upper limits to $P_l$. The crosses
  are the features that are affected by the interference described in
  \S.\ref{obs} and which were excluded from our polarization
  analysis. The open circle is a feature that likely does not belong
  to the filament and which has been excluded in the subsequent
  analysis. The solid line is a best fit relation between the maser
  velocity and position off-set.}
\label{Fig:vel2}
\end{figure}

As can be seen in Figs.~\ref{Fig:FULL} and \ref{Fig:MOSAIC} the \water
masers around Cepheus~A~HW2 show a large variety of structures. In our
observations several of the maser structures found in T01b and G03
were not detected, even though our sensitivity is within a factor of 2
of those of T01b ($\sim$6~mJy) and G03 ($\sim$25~mJy). We did
detected a strong linear maser structure in Field II that was not observed in
the previous observations. The changes in observed morphology are likely due to the rapid
variability of the \water masers of Cepheus~A, which show variations
on timescales as short as a few days \citep{Rowland86}. Here we
discuss the masers of the 4 distinct regions detected in our
observations.

 {\it Field I:} The \water masers in this field show the most complex
 structure. The masers are located $\approx$150~AU on the sky south
 of the continuum source HW2 and have been previously detected in
 T96, T01b and G03. In T01b this region was named R4
 and it was
 proposed that the masers of the NW corner (named R4-A) originate in
 a bow-shock structure produced by the wind of an undetected protostar
 near R4-A. The features in the SE could be connected to R4-A
 and produced by a shock moving $\sim$4--7~\kms\ to the NE. In G03,
 where these masers were observed with MERLIN, the masers are, instead
 of in a bowshock, hypothesized to occur in an
 expanding shockwave in a rotating proto-stellar disk enclosing a
 central mass of
 $\sim$3~$M_\odot$. Here we only detect several bright features
 making up an incomplete disk with $V_{\rm lsr}\approx -4.0$, $-1.5$ and
 $-19.0$~\kms. We fitted our maser feature positions to the disk proposed by
 G03 using a flux density weighted least square method. Keeping the
 inclination and position angle fixed with the G03 model values
 ($50^\circ$ and $142^\circ$ respectively) a fit was made for the Right
 Ascension and Declination offset of the disk center and for the disk Radius
 ($R_d$). The result is
 shown in Fig.~\ref{Fig:shell}a. Our disk center offset position is only
 $\sim$7~mas SW of the position determined by G03 while the error on
 the reference position determination in this paper combined with that
 of G03 is estimated to be $\sim$27~mas. Our fitted disk radius 
 $R_d=34$~mas. Considering we only detect a small part of the disk and since
 the masers in the SE corner are spread over a large area we estimate
 the systematic error in our radius determination to be $4$~mas,
 larger than the formal fitting uncertainty of $\sim$1~mas. Comparing
 $R_d$ with the radius determined at Epoch 2000.27 by G03
 ($R_d=38.1\pm0.1$~mas) we find that
 the disk has not expanded in the 4 years after their observations. It
 possibly even decreased in radius. In G03 it was found that the
 expansion velocity decreased from 30--40~\kms\ in 1996 to
 $\sim$13~\kms\ in 2000. This strong deceleration apparently has
 continued and may be due to mass loading of the disk as matter
 is swept up during the expanding shockwave. As a result a stationary
 shock may
 have formed where the circumstellar outflow collides with the much
 denser surrounding medium. This could also explain the disappearance of
 the brightest disk masers observed in G03, since for higher shock number
 densities ($\gtrsim~10^{12}$~cm$^{-3}$) the masers will be quenched. 


\begin{figure*}
   \resizebox{\hsize}{!}{\includegraphics{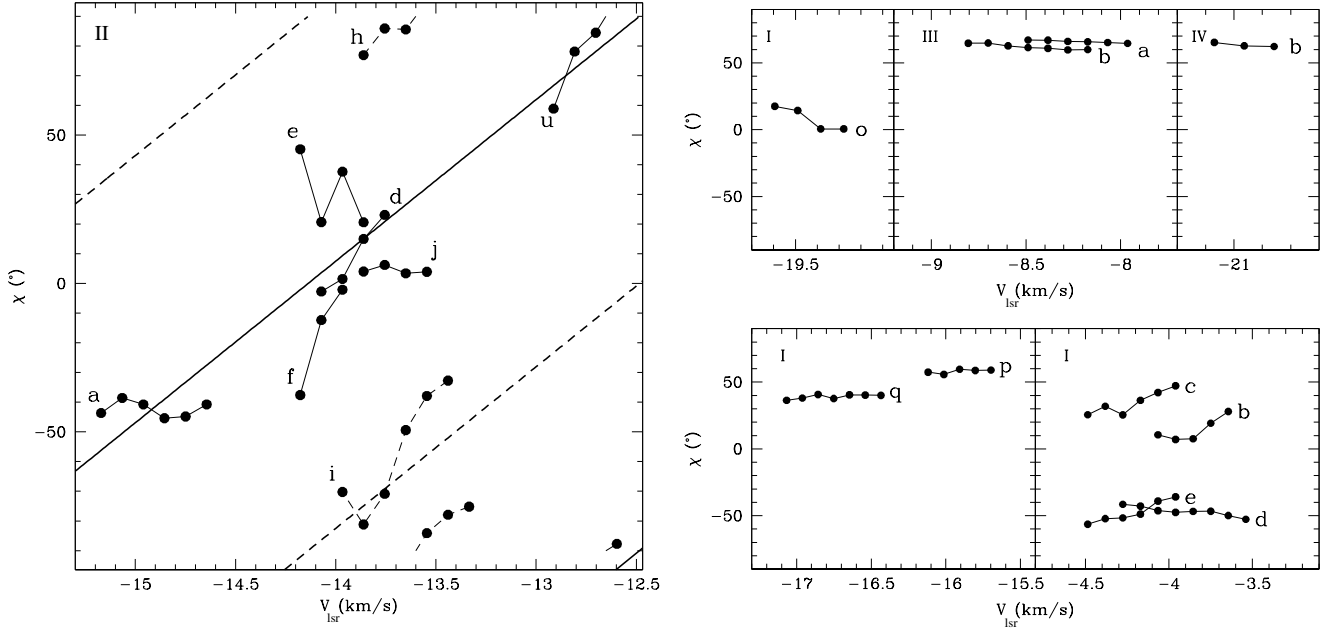}}
   \hfill
\caption[EVPA2]{(Left) $\chi$ for the \water maser filament
  features in Field II that have measured linear polarization
  vs. their velocity. The connected dots belong to the individual
  features which are labeled with the feature identifier. We excluded the
  seemingly unrelated feature denoted by the open circle in
  Fig.\ref{Fig:vel2}. The thick solid lines (with a 180$^\circ$ ambiguity) is a linear fit to the change in $\chi$
  along the feature. The thick dashed lines indicates the linear fit when including the 90$^\circ$ flip in
  $\chi$ with respect to the magnetic field direction when the magnetic field angle
  to the line-of-sight $\theta$ becomes larger than $\theta_{\rm
    crit}\sim$55$^\circ$. For the features connected with the solid
  lines we expect the $\chi$ to be parallel to the direction of the
  magnetic field while for those connected with dashed lines, $\chi$ is
  perpendicular to the magnetic field. (Right) $\chi$ for the
  \water maser in Field I, III and IV that have measured linear
  polarization vs. their velocity. The connected dots belong to the
  individual features that are labeled with their corresponding
  identifier. The boxes are labeled with the field number.}
\label{Fig:EVPA2}
\end{figure*}

 {\it Field II:} The \water masers in Field II make up a newly
 discovered filamentary structure $\sim$690~AU East of HW2 at a
 position angle (PA) of $66.0^\circ \pm 0.2^\circ$ and with a length
 of $\sim$60~AU. This structure also nearly coincides with 12~GHz
 methanol masers (at different $V_{\rm lsr}$) located
 $\sim$40$\pm10$~mas to the NW, which show a linear structure with
 similar PA.  As seen in Fig.~\ref{Fig:vel2} there is a velocity
 gradient along the filament from $\sim\!-15$~\kms\ in the NE to
 $\sim\!-12.5$~\kms\ in the SW. The maser structure bears resemblance
 to the masers in R1, R2 and R3 of T01 found towards the West of HW2
 with a similar PA, although the masers in Field II are all
 blue-shifted with respect to the systematic velocity of HW2 while
 those in R1, R2 and R3 were red-shifted. The masers are too far East
 to be considered part of the rotating maser disk around HW2 which is
 thought to have a radius of 300~AU (T96). The elongated appearance of
 the Field II maser structure suggests a shocked nature as expected
 from maser theory \citep{Elitzur89}. Although it is located at a
 significant distance from HW2 we suggest that the maser structure is
 due to the interaction of the HW2 outflow with the circumstellar
 molecular cloud medium. Then, similarly as for the Western R1
 features in T01b, the velocity shift of $\sim$2.5~\kms~along the
 maser filament could be due to acceleration of maser gas by the YSO
 outflow. If the masers are indeed created by shocks induced by the
 HW2 outflow this would indicate that at $\sim$690~AU the outflow has
 an opening angle of $\sim$115$^\circ$, similar to the opening angle
 of $\sim$110$^\circ$ estimated for the R1 masers at 150~AU in T01.

 {\it Field III:} The 2 maser features detected in Field III,
 approximately 550~AU South of HW2 at $V_{\rm lsr}\approx$-8.5\kms,
 likely belong to the arc structure R5 described in T01a and
 T01b. These masers are thought to be part of a spherical shell
 surrounding a protostar that has possibly been identified in
 \citet{Curiel02}. We do not detect the long maser arc seen in T01a
 and T01b. While the brightest maser feature in our observations of
 Field III is $\simeq$20~Jy~beam$^{-1}$, the brightest maser features
 of R5 in T01a and T01b (separated by 8.5~yr with respect to our
 observations) had flux densities of
 $\simeq$200~Jy~beam$^{-1}$. However, the PA ($\sim$41$^\circ$) of the
 extended maser structure of our observations (seen in
 Fig.~\ref{Fig:LP3}) agrees with the direction of the maser
 curve. Fig.~\ref{Fig:shell}b shows the maser shell from T01a
 extrapolated in time using the updated shell parameters determined in
 G03. The curves indicate a shell expansion of $2.5\pm
 0.1$~mas~yr$^{-1}$ and a motion of the expansion center of $1.4\pm
 0.1$~mas~yr$^{-1}$ toward PA 126$^\circ$. The near-perfect alignment
 of the expanding shell with the maser features is remarkable, as we
 earlier estimated our reference position to be accurate to
 $\sim$25~mas. This likely indicates that we underestimated our
 positional accuracy and that the actual accuracy is better than
 $\sim$10~mas. In addition, our results indicate that the maser
 shell has been freely expanding during the past 8.5 years without any
 indication of deceleration.

 {\it Field IV:} The masers in Field IV are weak and are aligned at a
 PA $\sim\!-6^\circ$. They are located within $\sim$75~AU
 of HW2 at $V_{\rm lsr}\approx$-21.0\kms\ and are probably part of the
 rotating maser disk around HW2 proposed in T96.

\subsection{Linear Polarization}
\label{disclp}

Linear polarization is often affected by Faraday Rotation due to free
electrons along the line of sight through the interstellar
medium. However, the Faraday rotation induced in a typical molecular
cloud with fairly strong magnetic field (size $D\sim$0.1~pc, electron
density $n_e\sim$1~cm$^{-3}$ and $B_{||}\sim$1~mG) is only $\sim
0.9^\circ$ at 22.235~GHz. The rotation induced in the extreme condition of a highly
magnetized maser cloud (up to 1~G) is similar or less. As no compact
HII regions, in which Faraday rotation could be significant, are
located in front of the maser features, we can safely assume the measured
$\chi$ is not affected by Faraday rotation.


As discussed in Appendix~\ref{method}, the polarization vectors determined from
polarization observations of masers in a magnetic field are either
parallel or perpendicular to the magnetic field lines. Thus, the
polarization vectors contain information on the morphology of the
magnetic field but suffer from a $90^\circ$ degeneracy. The fractional
linear polarization depends on the maser saturation level as well as
the magnetic field angle $\theta$. Thus, we can use the measurements
of $P_l$ together with our model results for the saturation level
(through the emerging brightness temperatures) to lift the degeneracy
between the polarization vectors and the direction of the magnetic
field for several of our maser features.

The polarization vectors observed in the maser fields around
Cepheus~A~HW2 are shown in Fig.~\ref{Fig:MOSAIC} while $P_l$ is listed
in Table~\ref{Table:results}. The strongest linear polarization
$P_l\approx~11$\% was found in Field III. This is consistent with the
fact that the brightness temperature analysis concluded that the
masers in this field are saturated. Using the brightness temperature
determined from the models, adjusted for the difference in maser decay
and cross-relaxation rate as described in Appendix~\ref{method}, we find using Fig.~\ref{Fig:lintheta}, that for the masers in Field III, $65^\circ < \theta <
70^\circ$. As was shown in \citet[][~hereafter V05]{Vlemmings05c} this
is the magnetic field angle in the unsaturated (or least saturated)
maser core. Thus, as $\theta>\theta_{\rm crit}$, the magnetic field
direction is perpendicular to the polarization vectors. As can be seen
in Fig.~\ref{Fig:shell}b this means the magnetic field, at
PA$\sim$155$^\circ$, is perpendicular to the expanding shell found in
T01a and thus radial from the central embedded proto-star.

In Field IV fairly strong linear polarization was detected in one of
the weak maser features. As the brightness temperature of these masers
is relatively low and they are unlikely to be saturated, $P_l=1.35\%$
indicates that the magnetic field angle $\theta>70^\circ$. Thus also
in Field IV the magnetic field direction is perpendicular to the
polarization vector with a PA$\sim$154$^\circ$, more or less along the
large scale maser disk proposed in T96 with a PA$\sim$135$^\circ$ and
radial toward HW2.

The \water masers in the circumstellar disk of Field I are found to
have $P_l<1\%$, consistent with their being only slightly saturated. With
$T_{\rm b}$ determined earlier, we find, again using
Fig.~\ref{Fig:lintheta}, that $\theta$ is either close to $\theta_{\rm
  crit}$ or $\theta<25^\circ$. As seen in Fig.~\ref{Fig:shell}a, the
polarization vectors mostly lie along the disk curvature for most
features except I.c and I.d, where we could be seeing a $90^\circ$
flip. If we assume that for all features, except I.c and I.d,
$\theta\lesssim\theta_{\rm crit}$, the magnetic field in Field I lies
along the disk. However, if most features have
$\theta\gtrsim\theta_{\rm crit}$ except for I.c and I.d, the magnetic
field is radial in the \water maser region of the rotating disk.
Additionally, as in either case, $\theta$ is close to $\theta_{\rm
  crit}=55^\circ$ the magnetic field angle with respect to the line-of-sight is
similar to the disk inclination axis, which would imply the magnetic field lies in the plane of the disk. However, as seen in
Table~\ref{Table:results}, the magnetic field direction changes
between the neighboring maser features making a large scale alignment
unlikely.

Now we show that the polarization characteristics of the masers in
Field II are consistent with the interaction between a radial magnetic
field in the outflow of HW2 and a magnetic field perpendicular to the
Galactic plane in the surrounding molecular cloud. The fractional
polarization of the maser in the filamentary structure of Field II is
on average slightly less than that in Field I. This is expected since
the masers in Field II were found to be unsaturated. The high
polarization of feature II.h likely indicates that there $\theta$ is
close to $90^\circ$. As seen in Fig.~\ref{Fig:MOSAIC}, there is
evidence of a gradient in polarization angles along the maser
filament. In addition to the gradient along the maser filament, we see
in the left panel of Fig.~\ref{Fig:EVPA2} that the polarization angle
$\chi$ rotates across individual maser features, similar to that seen
in the cocoon masers of \citet{Leppanen98}. Such rotation of $\chi$ is
not observed for any of the maser features of the other fields shown
in the right panels of Fig.~\ref{Fig:EVPA2}. The variation of $\chi$
with velocity can be described with a linear gradient, using an flux
density weighted least square method allowing for the $90^\circ$ flip
in $\chi$ that occurs when $\theta>\theta_{\rm crit}$.  We find that
$\chi$ increases linearly from $\sim\!-50^\circ$ on the maser feature
in the NE to $\sim$90$^\circ$ on the feature in the SW. This implies
that II.h and II.i undergo the $90^\circ$ flip which was already
expected for II.h due to its high polarization.

\begin{figure}
   \resizebox{\hsize}{!}{\includegraphics{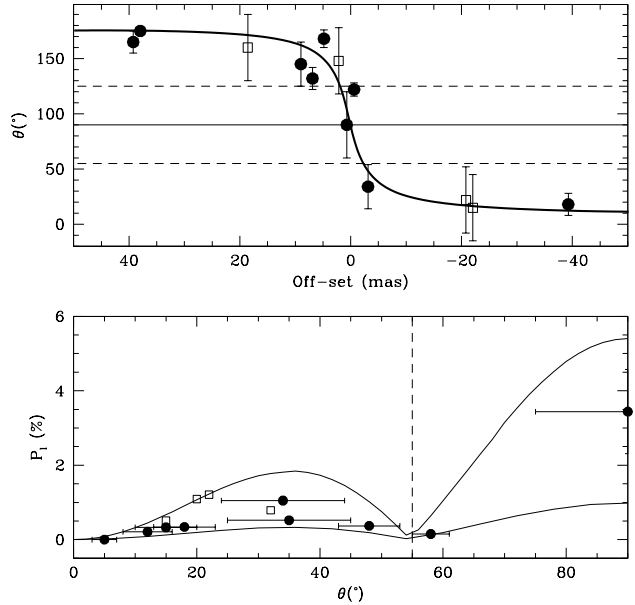}}
   \hfill
\caption[ANGMOD]{(bottom) The magnetic field angle $\theta$ estimated from the fractional polarization $P_l$ measurement. The vertical dashed line indicates $\theta_{\rm crit}$. The solid dots are the measured $P_l$ of the Field II maser features and the open squares are upper limits. The error bars in $\theta$ are determined from the allowed range of $\theta$ in the $P_l$ vs. $\theta$ models of V05 shown for masers with emerging brightness temperatures $T_{\rm b}\Delta\Omega=10^9$ and $10^{10}$K~sr. (lower and upper solid line respectively). (top) A model for the change of magnetic field angle $\theta$ along the \water maser filament. The thick solid line is the proposed model. The solid horizontal line indicates where the magnetic field direction changes. Between the dashed horizontal lines, which denote $\theta_{\rm crit}$, the polarization vectors are perpendicular to the direction of the magnetic field. Above and below $\theta_{\rm crit}$ the polarization vectors are parallel to the magnetic field.}
\label{Fig:THETAMOD}
\end{figure}

Similar to the model for the variation in polarization angle $\chi$,
we have constructed a model for the variation of the angle between the
magnetic field direction and maser propagation axis $\theta$ along the
maser filament. The model, shown in Fig.~\ref{Fig:THETAMOD}, is fully
consistent with the maser brightness temperatures and the fractional
linear polarizations as well as the inferred $90^\circ$ flip of
polarization angle. We have determined $\theta$ and its error bars
from the relation between $P_l$ and $\theta$ for unsaturated masers
shown in Fig.~\ref{Fig:lintheta}. (Note that to the accomodate the
direction change of the magnetic field, the model $\theta$ ranges from
$0^\circ$ to $180^\circ$, with the direction change occurring at
$\theta=90^\circ$). We find that the polarization measurements are
consistent with an initial slow change in $\theta$ until halfway along
the filament the magnetic field changes sign over
$\approx$10~mas.

Combining the $\chi$ and $\theta$ variation models we thus find that
at the NE corner of the filament the magnetic field is pointing toward
us with $\theta\sim$10$^\circ$ and PA$\sim-50^\circ$, while in the SW
corner it is pointing away from us with $\theta\sim$20$^\circ$ and
PA$\sim$90$^\circ$. We interpret this change of the magnetic field as
being due to the interaction of the magnetic field related to the HW2
YSO outflow and that related to the surrounding medium. Although the
magnetic field structure in the Cepheus~A region is complex,
\citet{Jones04} find using Near Infrared imaging polarimetry, that the
large scale field threading Cepheus~A is almost perpendicular to the
galactic plane with PA$=125^\circ$. This corresponds to
PA=$-55^\circ$, consistent with $\chi=-50^\circ$ found in the NE
corner of the maser filament. \citet{Jones04} also argue that the
magnetic field in the HW2 outflow is radial with respect to HW2. This
implies an angle of $115^\circ$ at the location of Field II, very
consistent with the PA in the SW corner of the \water maser filament,
especially as we are probably not probing the full polarization vector
rotation along the filament.

\subsection{The Magnetic Field}

\subsubsection{The Magnetic Field Strength}

The magnetic field strength was determined from circular polarization
measurements for 14 maser features in Field I, II and III. In Field I the magnetic
field strength varies from $B=-290$ to $527$~mG, while in the other
fields we find $B$ between $-54$ and $128$~mG. While these latter
magnetic field strengths are comparable to the typical field strength
determined from other \water maser observations ($10$--$100$~mG), the
field strengths determined in Field I are several times
higher. However, previous observations were typically performed using
single-dish \citep[e.g.][]{Fiebig89} or lower resolution interferometers \citep[e.g][]{Sarma02},
and due to blending of a large
number of maser features, the magnetic field strength determined with
single-dish and VLA observations could be more than a factor of 2
smaller than the actual field strengths \citep{Sarma01}. The only other \water maser magnetic
field strength for Cepheus~A was determined by \citet{Sarma02} with
the VLA, who found $B=$-3.2~mG for a feature located more
than 2~arcsec East of our observed maser region. 

An additional complication to the accurate determination of magnetic
field strengths is the occurrence of velocity gradients along the maser
amplification path. This was investigated in V05, where it was found
that for velocity gradients of $\sim$1.5~\kms\ along the maser, the
magnetic field could be underestimated by $\sim$40\%. From a total
intensity line profile analysis as described in \citet{Vlemmings05a} we estimate the
typical velocity gradient for our masers to be $\sim$1~\kms. For
partly or fully saturated masers with $\Delta v_{\rm th}>1.5$~\kms, V05 finds
that the magnetic field is overestimated by not more than a few
percent. However, for the unsaturated masers in Field II the field
strengths have most likely been underestimated by $\sim$30\%.

The magnetic field dependence on $\theta$ introduces further
uncertainties. While for low brightness temperature masers $B$ is
straightforwardly dependent on $\cos\theta$, this relation breaks down
for higher brightness temperatures. This was first investigated in
\citet{Nedoluha92} and later shown in more detail in \citet{WW01} for masing
involving low angular momentum transitions. The specific case for the
22~GHz \water masers was again shown in V02 and their figure 7 is
reproduced here as Fig.~\ref{Fig:BHETA}.  In the figure we see that
for increasing saturation there is a large range of $\theta$ where the
magnetic field is actually overestimated. As we have been able to
estimate $\theta$ for several of the observed masers we can also
estimate the influence on the magnetic field strength. For the masers
in Field I $\theta$ is thought to be close to $\theta_{\rm crit}$,
while the masers are saturating. This means the actual magnetic field
is approximately 10\% higher than the measured field strength. Thus we
estimate the average field in Field I to be $\sim$250~mG with a
maximum of $\sim$650~mG. The masers in Field II are however mostly
unsaturated and as a result $|B|=B_{||}/\cos\theta$. As we only detected a
magnetic field strength at the edges of the filament, where we
estimated $\theta$ to be between 10$^\circ$ and 20$^\circ$, the field
in the NE of the filament is $\sim$55~mG pointing toward us while it
is $\sim$70~mG and pointing away from us in the SW. Finally, the
masers in Field III were found to have $65^\circ<\theta<70^\circ$. As
they are saturated the magnetic field strength is likely $\sim$20\%
less than determined from our fits, indicating that $|B|=30$--100~mG.

Aside from the large magnetic field strength, the maser structure in
Field I is also characterized by field reversals on small scales. The
magnetic field is found to reverse over less than 0.1~mas, which
corresponds to $\sim\!10^{12}$~cm. This argues against a large scale
alignment of the magnetic field with the maser disk. The magnetic
field is likely frozen into high density maser clumps in a turbulent
medium.   If the masers exist in a shocked
region where the magnetic pressure supports the cloud and dominates
the gas pressure higher magnetic fields can be obtained. Using formula
4.5 from \citet{Kaufman96}, 
\begin{equation}
B\sim80({{n_0}\over{10^8~{\rm cm}^{-3}}})^{1/2}({{v_s}\over{10~{\rm
km~s}^{-1}}})~{\rm mG}, 
\label{eqb}
\end{equation} 
where $n_0$ is the pre-shock H$_2$ density and $v_s$ is the shock velocity, we find that for a shock
velocity $v_s=10$~\kms\ as estimated for Field I, a magnetic field
$B=600$~mG can be reached if the pre-shock number density
$n_0=5.6\times10^9$~cm$^{-1}$. Estimating the pre-shock magnetic field
using 
the empirical relation of \citet{Crutcher99} $B\propto n^{0.47}$ from
the density and magnetic field found at the edge of NH$_3$ molecular
clouds ($B=0.3$~mG, $n=2\times10^4$~cm$^{-3}$; \citealp{Garay96})
yields $B_0\approx$100~mG, which is almost 2 orders of magnitude
larger than the typical pre-shock magnetic field strength
($\sim$1~mG).  Also, when determining the number density of hydrogen in the
shocked \water maser region using the relation from
\citet{Crutcher99}, the magnetic fields imply densities
$n_{H_2}=5\times10^9$--$2\times10^{11}$~cm$^{-3}$.  While the low-end values
for $n$ are reasonable for \water masers, the high end
($>10^{10}$~cm$^{-3}$) is unlikely, as such high densities quench the
maser population inversion. Thus, the magnetic field strength in the pre-shock
medium of the proto-stellar disk is likely enhanced by the pressence
of a nearby magnetic dynamo.

Using Eq.~\ref{eqb} to estimate the pre-shock number density near the
maser filament in Field II yields, assuming $v_S=13$~\kms\ similar to
the shock velocity in R1 to R3 of T01b, $n_0=3\times 10^7$--$1\times
10^8$~cm$^{-3}$. Scaling with $B\propto n^{0.47}$ this implies, for
the pre-shock magnetic field $B_0\approx$10--15~mG, similar to the
magnetic field determined for comparable densities in the OH masers of
Cepheus~A \citep{Bartkiewicz05}. For the number density in the shocked
region this implies $n_{H_2}=3.5\times 10^8$--$4.7\times 10^9$~cm$^{-3}$,
typical for \water masers.


\subsubsection{The influence of the Magnetic Field}

We now examine the influence of the magnetic field on the molecular
outflow around HW2. When the magnetic field pressure becomes equal to
the dynamic pressure in the outflow the magnetic field will be able to
influence or even control the molecular outflow. Defining $B_{\rm
crit}$ the critical magnetic field where the dynamic and magnetic
pressure are equal, we find \begin{equation} B_{\rm crit}=(8\pi\rho
v^2)^{1/2}, \end{equation} where $\rho$ and $v$ are the density and
velocity of the maser medium respectively. Assuming an outflow
velocity of $\sim$13~\kms\ we find $B_{\rm crit}\approx$30, 100 and
350~mG for number densities of $n_{H_2}=10^8, 10^9$ and
$10^{10}$~cm$^{-3}$ respectively. This means that in all the \water
maser regions where we measured the magnetic field strength the
magnetic pressure is approximately equal to the dynamic pressure, as
was previously found in \citet{Sarma02}. As OH maser polarization
observations indicate that this also holds in the lower density
pre-shock regions, we conclude that the magnetic field strength is
capable of controlling the outflow dynamics.

\section{Summary}
\label{sum}

Using polarimetric VLBA observations of the \water masers around
Cepheus~A~HW2 we have been able to measure the magnetic field strength and direction
in great detail at sub-AU scales. We detected \water masers over an area of $\sim$1~$\arcsec$ in 4
distinct fields. For each of the fields we derived physical properties
and several intrinsic properties of the masers.

{\it Field I:} The \water masers in this field occur in what was
proposed in G03 to be a spherical shockwave expanding through a
circumstellar disk. We find that between the G03 MERLIN observations
in 2000 and our observations the maser ring has not expanded and
conclude that the expanding shockwave has been severely decelerated,
possibly due to mass-loading. From our maser models and the measured
brightness temperatures we find that the typical maser beaming angle
in this field is $\sim$5$\times 10^{-2}$~sr implying maser
amplification lengths of several AU. The masers are approaching
saturation. The magnetic field strength is strong (on average $250$~mG
and as high as $650$~mG) and shows direction reversal on scales of
$\sim$10$^{12}$~cm. This can be due to the fact that the magnetic
field is frozen into a dense and turbulent medium although the linear
polarization vectors indicating the magnetic field direction follow
the disk and the magnetic field angle with respect to the line-of-sight $\theta$
is approximately equal to the disk inclination. The high magnetic
field strengths indicate that the field is enhanced by a nearby
magnetic dynamo.

{\it Field II:} This field consists of a newly discovered maser
filament at $\sim$690~AU East of Cepheus~A~HW2 with a
PA$=66^\circ$. It is likely the result of the shock interaction
between the HW2 outflow and the surrouding molecular cloud and implies
a large opening angle ($115^\circ$) of the outflow. The maser beaming
angle in Field II is $\sim$10$^{-2}$~sr with maser path lengths of
$\sim$2~AU while the masers are unsaturated. We find a clear velocity
and magnetic field orientation gradient along the filament consistent
with the interaction between a radial magnetic field in the HW2
outflow and the magnetic field in the surrounding Cepheus~A complex
which is almost perpendicular to the Galactic plane. The magnetic
field strength of $50$--$70$~mG is typical for \water masers found in SFRs.

{\it Field III:} The masers of Field III make up a small part of the
shell structure found in T01 and, even though our maser reference
position has an estimated error of up to $25$~mas, are fully
consistent with the shell expansion model parameters estimated in
G03. We find a magnetic field strength between 30--100~mG, consistent
with other SFR \water maser polarization measurements and find that
the magnetic field direction is along the shell expansion direction,
radial from the central embedded proto-star. As these maser have the highest
measured linear polarization, $P_l=10\%$, we can conclude that they
are saturated. The beaming angle is consistent with a spherical maser
geometry.

{\it Field IV:} Located close to HW2, the maser in this field are weak
and no magnetic field strength was determined. The upper limits of
$B_{||}\approx$~500~mG. The linear polarization indicates
that the magnetic field is either aligned with the \water maser disk
around HW2 or radial toward HW2.

\section{Conclusions}
\label{concl}

Strong magnetic fields of up to $\sim$600~mG have been measured in the
\water masers around Cepheus~A~HW2. The strongest magnetic fields were
measured in the maser structure that was identified as a circumstellar
disk (G03), suggesting the nearby presence of a dynamo source. The
field strengths determined in the maser regions further from the
central source HW2 are 30--100~mG, consistent with earlier
VLA, VLBA and single dish measurements of SFRs. The high magnetic
field strengths indicate that the magnetic pressure is similar to the
dynamic pressure in the outflows around HW2. Thus, the magnetic fields
likely play a large role in supporting the molecular cloud and shaping the outflows in this very active high-mass star-forming region.

\begin{acknowledgements}
WV thanks R.J. Cohen for usefull discussions
and comments.  WV was supported by an EC Marie Curie Fellowship under
contract number MEIF-CT-2005-010393. JMT acknowledges partial financial support from MCYT (Spain) grant AYA2005-08523-C03-02. 
\end{acknowledgements}

\appendix

\section{Polarization Modeling and Analysis}
\label{method}

Here we describe the modeling and analysis of the 22~GHz \water maser linear
and circular polarization used in this paper to determine the magnetic
field strength, saturation level and intrinsic thermal line width of
the maser features.

\begin{figure}
   \resizebox{\hsize}{!}{\includegraphics{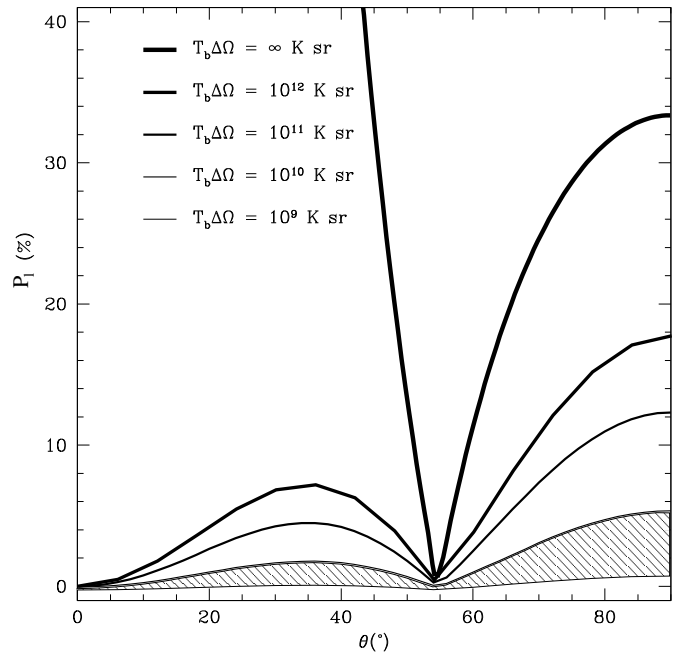}}
   \hfill
\caption[Q-Theta]{The angle $\theta$ between the maser propagation
  direction and the magnetic field vs. the fractional linear
  polarization $P_l$ for different values of emerging maser brightness
  (for $[\Gamma+\Gamma_\nu]=1$~s$^{-1}$). The thick solid line denotes
  the theoretical limit from \citet{Goldreich73} for a fully saturated
  maser. The shaded area is the region of emerging brightness
  temperatures found for the masers in Field II.}
\label{Fig:lintheta}
\end{figure}

\begin{figure}
   \resizebox{\hsize}{!}{\includegraphics{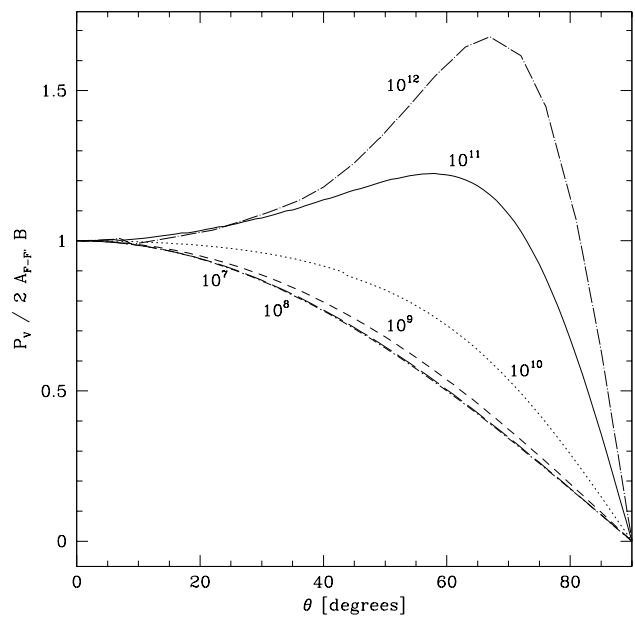}}
   \hfill
\caption[B-Theta]{$\theta$-dependence of Eq.~\ref{eq2} for increasing
   emerging brightness temperature $T_{\rm
   b}\Delta\Omega$ (for $[\Gamma+\Gamma_\nu]=1$~s$^{-1}$) from V02. The lines
   for $T_{\rm b}\Delta\Omega = 10^{7}$ and $10^8$ coincide and are
   the same as the lines for lower brightness temperatures. For fully unsaturated masers the dependence is equal to $\cos\theta$}
\label{Fig:BHETA}
\end{figure}

\subsection{Circular Polarization}

For the analysis of the circular polarization spectra we used the full
radiative transfer non-LTE interpretation, which was thoroughly
described in V02. There the coupled equations of state for the 99
magnetic substates of the three dominant hyperfine components from
\citet{Nedoluha92}~(hereafter NW92) were solved for a linear maser in
the presence of a magnetic field. The emerging maser flux densities of the
resulting spectra are expressed in $T_{\rm b}\Delta\Omega$, where
$T_{\rm b}$ is the brightness temperature and $\Delta\Omega$ is the
beaming solid angle. It was found in NW92 that the emerging brightness
temperature scaled linearly with $(\Gamma + \Gamma_\nu)$, which
are the maser decay rate $\Gamma$ and cross-relaxation rate
$\Gamma_\nu$. For the 22~GHz \water masers, $\Gamma$ is typically
assumed to be $\lesssim 1$~s$^{-1}$. In star-forming regions it has been
found that $\Gamma_\nu\!\approx$2~s$^{-1}$ for $T\sim$400~K
and $\Gamma_\nu\!\approx$5~s$^{-1}$ for $T\sim$1000~K
\citep{Anderson93} and thus the models from V02 (where $(\Gamma +
\Gamma_\nu)\!=$1~s$^{-1}$) have been adjusted to these values.

The model results further depend on the intrinsic thermal line-width
$\Delta v_{\rm th}$ in the maser region, where $\Delta v_{\rm th} \approx
0.5(T/100)^{1/2}$ with $T$ the temperature of the masing gas.  Model
spectra for a grid of $\Delta v_{\rm th}$ between 0.8 and 2.5~\kms,
corresponding to temperatures between 250 and 2500~K, were directly
fitted to the observed I and V spectra using a least square fitting
routine. As described in V02 the spectral fitting for the non-LTE
analysis requires the removal of the scaled down total power spectrum
from the V-spectrum to correct for small residual gain errors between
the right- and left-polarized antenna feeds. This was typically found
to be $\sim 0.5\%$ of the total power. The best fit model thus
produced the line of sight magnetic field $B_{||}$ and the thermal
line-width $\Delta v_{\rm th}$ as well as the maser emerging brightness
temperature $T_{\rm b}\Delta\Omega$. However, the uncertainties in
$\Delta v_{\rm th}$ and $T_{\rm b}\Delta\Omega$ are large, as they are
strongly affected by maser velocity gradients (\citealp{Vlemmings05a}; V05). Additionally, $T_{\rm b}\Delta\Omega$ depends on the actual value
of $(\Gamma+\Gamma_\nu)$. We estimate the uncertainties in the fit for
$\Delta v_{\rm th}$ to be $\sim$0.3~\kms\ and the uncertainty in $\log(T_{\rm
b}\Delta\Omega)$ to be $\sim$0.4. As the magnetic field depends on the
intrinsic thermal line-width and emerging brightness temperature
(V02), this leads to an added uncertainty in the magnetic field
determination of $\sim$15\% which has been included in the
formal fitting errors.

When a direct model fit was not possible, we used
the relation between the
magnetic field strength $B$ and percentage of circular polarization
$P_V$.
\begin{eqnarray}
P_{\rm V} & = & (V_{\rm max} - V_{\rm min})/I_{\rm max} \nonumber\\
& = & 2\cdot A_{F-F'}\cdot B_{\rm [Gauss]} \rm{cos}\theta/\Delta v_{\rm L}[\rm{k
m~s^{-1}}].
\label{eq2}
\end{eqnarray}
Here $\theta$ is the angle between the maser propagation direction and
the magnetic field ($0^\circ<\theta<90^\circ$) and $\Delta v_{\rm L}$
is the maser full width half maximum (FWHM). $V_{\rm max}$ and $V_{\rm
  min}$ are the maximum and minimum of the circular polarization and
$I_{\rm max}$ is the maximum total intensity maser flux density. The
coefficient $A_{\rm F-F'}$ describes the relation between the circular
polarization and the magnetic field strength for a transition between
a high ($F$) and low ($F'$) rotational energy level. $A_{\rm F-F'}$
depends on $\Delta v_{\rm th}$ and maser saturation level as described
in NW92 and V02 as well as on velocity and magnetic field gradients
along the maser path as shown in V05. We used $A_{\rm F-F'} = 0.012$,
which is the typical value we found for the maser of
Cepheus~A~HW2. For maser features where no circular polarization was
detected the $3\sigma$ upper limits were determined using
Eq.~\ref{eq2} with $P_V=6\sigma_V / I_{\rm max}$, with $\sigma_V$
being the rms noise on the maser V-spectrum determined after Hanning
smoothing the spectrum ($\sigma_V\sim$5--8~mJy).

For maser brightness temperatures $T_{\rm b}\Delta\Omega > 10^9$~K~sr
it was shown in NW92 that the ${\rm cos}\theta$ dependence of
Eq.~\ref{eq2} breaks down introducing a more complex dependence on
$\theta$. This was later shown in more detail in \citet{WW01} for
masing involving angular momentum $J=$1--0 and $J=$2--1
transitions. In V02, figure 7 shows the derived magnetic field
strength dependence on $\theta$ to $A_{F-F'}$ for the 22~GHz $J=$6--5
transition. This figure is repeated here as Fig.~\ref{Fig:BHETA}.

\subsection{Linear Polarization}

Maser theory has shown that the percentage of linear polarization
$P_l$ of \water masers depends on the degree of saturation and the
angle $\theta$ between the maser propagation direction and the
magnetic field \citep[e.g. NW92; ][]{Deguchi90}. Figure 7 of NW92 shows
the relationship between $\theta$ and $P_l$ while their Figure 8 shows
the $P_l$ dependence on saturation level. Fig.~\ref{Fig:lintheta}
shows the dependence of $P_l$ on $\theta$ for various emerging
brightness temperatures as calculated from the models of NW92 and
V02. A high linear polarization fraction ($P_l>5\%$) can only be
produced when the maser is saturated.  Additionally, the polarization
vectors are either perpendicular or parallel to the magnetic field
lines, depending on $\theta$. When $\theta>\theta_{\rm
  crit}\approx$55$^\circ$ the polarization vectors are perpendicular
to the magnetic field, and when $\theta<\theta_{crit}$ they are
parallel \citep{Goldreich73}. Consequently, the linear polarization
vectors can flip $90^\circ$ at very small scales as was observed in
for instance circumstellar SiO masers \citep{Kemball97}.

\end{document}